\documentclass[twocolumn,prl,showpacs,superscriptaddress,floatfix]{revtex4}
\usepackage{graphicx}
\usepackage{amsmath}
\newcommand{\Cl}{Sr$_2$CoO$_3$Cl}
\begin{document}
\title{A different look at the spin state of Co$^{3+}$ ions in CoO$_{5}$ pyramidal coordination}
\author{Z. Hu}
\affiliation{II. Physikalisches Institut, Universit\"at zu K\"oln, Z\"ulpicher
Str. 77, 50937 K\"oln, Germany}
\author{Hua Wu}
\affiliation{Max-Planck-Institut f\"ur Physik komplexer Systeme, N\"othnitzer
Str. 38, 01187 Dresden, Germany}
\author{M. W. Haverkort}
\affiliation{II. Physikalisches Institut, Universit\"at zu K\"oln, Z\"ulpicher
Str. 77, 50937 K\"oln, Germany}
\author{H. H. Hsieh}
\affiliation{National Synchrotron Radiation Research Center, 101 Hsin-Ann Rd.,
Hsinchu 30077, Taiwan}
\author{H. -J. Lin}
\affiliation{National Synchrotron Radiation Research Center, 101 Hsin-Ann Rd.,
Hsinchu 30077, Taiwan}
\author{T. Lorenz}
\affiliation{II. Physikalisches Institut, Universit\"at zu K\"oln, Z\"ulpicher
Str. 77, 50937 K\"oln, Germany}
\author{J. Baier}
\affiliation{II. Physikalisches Institut, Universit\"at zu K\"oln, Z\"ulpicher
Str. 77, 50937 K\"oln, Germany}
\author{A. Reichl}
\affiliation{II. Physikalisches Institut, Universit\"at zu K\"oln, Z\"ulpicher
Str. 77, 50937 K\"oln, Germany}
\author{I. Bonn}
\affiliation{Johannes Gutenberg-Universit\"at Mainz, Becher Wege 24, 55099
Mainz, Germany}
\author{C. Felser}
\affiliation{Johannes Gutenberg-Universit\"at Mainz, Becher Wege 24, 55099
Mainz, Germany}
\author{A. Tanaka }
\affiliation{Department of Quantum Matters, ADSM, Hiroshima University,
Higashi-Hiroshima 739-8530, Japan}
\author{C. T.\  Chen}
\affiliation{National Synchrotron Radiation Research Center, 101 Hsin-Ann Rd.,
Hsinchu 30077, Taiwan}
\author{L. H. Tjeng}
\altaffiliation{corresponding author} \affiliation{II. Physikalisches Institut,
Universit\"at zu K\"oln, Z\"ulpicher Str. 77, 50937 K\"oln, Germany}

\date{\today}

\begin{abstract}
Using soft-x-ray absorption spectroscopy at the Co-$L_{2,3}$ and O-$K$ edges,
we demonstrate that the Co$^{3+}$ ions with the CoO$_{5}$ pyramidal
coordination in the layered {\Cl} compound are unambiguously in the high spin
state. Our result questions the reliability of the spin state assignments made
so far for the recently synthesized layered cobalt perovskites, and calls for a
re-examination of the modeling for the complex and fascinating properties of
these new materials.
\end{abstract}


\maketitle

The class of cobalt-oxide based materials has attracted considerable interest
in the last decade because of expectations that spectacular properties may be
found similar to those in the manganites and cuprates. Indeed, giant magneto
resistance effects have been observed in the La$_{1-x}$$A$$_{x}$CoO$_{3}$
($A$=Ca,Sr,Ba) perovskites \cite{Briceno95} and $R$BaCo$_{2}$O$_{5+x}$
($R$=Eu,Gd) layered perovskites \cite{Martin97,Maignan99}. Very recently, also
superconductivity has been found in the Na$_{x}$CoO$_{2}$.$y$H$_{2}$O material
\cite{Takada03}. In fact, numerous one-, two-, and three-dimensional cobalt
oxide materials have been synthesized or rediscovered in the last 5 years, with
properties that include metal-insulator and ferro-ferri-antiferro-magnetic
transitions with various forms of charge, orbital and spin ordering
\cite{Fjellvag96,Aasland97,Kageyama97,Yamaura99,Loureiro00,Loureiro01,
Vogt00,Suard00,Fauth01,Burley03,Mitchell03,Moritomo00,Respaud01,Kusuya01,
Frontera02,Fauth02,Taskin03,Soda03}.

A key aspect of cobalt oxides that distinguish them clearly from the manganese
and copper materials, is the spin state degree of freedom of the Co$^{3+/III}$
ions: it can be low spin (LS, $S$=0), high spin (HS, $S$=2) and even
intermediate spin (IS, $S$=1) \cite{Sugano}. This aspect comes on top of the
orbital, spin (up/down) and charge degrees of freedom that already make the
manganite and cuprate systems so exciting. It is, however, also precisely this
aspect that causes considerable debate in the literature. For the classic
LaCoO$_{3}$ compound, for instance, various early studies attributed the low
temperature spin state change to be of LS-HS nature \cite{Goodenough65}, while
studies in the last decade put a lot of effort to propose a LS-IS scenario
instead \cite{Korotin96,Saitoh97}. More topical, confusion has arisen about the
Co spin state in the newly synthesized layered cobalt perovskites
\cite{Martin97,Maignan99,Yamaura99,Loureiro00,Loureiro01,
Vogt00,Suard00,Fauth01,Burley03,Mitchell03,Moritomo00,Respaud01,Kusuya01,
Frontera02,Fauth02,Taskin03,Soda03}. In fact, all possible spin states have
been claimed for each of the different Co sites present.  There is even no
consensus in the predictions from band structure calculations
\cite{Wu00,Kwon00,Wang01}.

In this manuscript we are questioning the reliability of the spin states as
obtained from magnetic, neutron and x-ray diffraction measurements for the
newly synthesized layered cobalt perovskites
\cite{Martin97,Maignan99,Yamaura99,Loureiro00,Loureiro01,
Vogt00,Suard00,Fauth01,Burley03,Mitchell03,Moritomo00,Respaud01,Kusuya01,
Frontera02,Fauth02,Taskin03,Soda03}.  We carried out a test experiment using a
relatively simple model compound, namely {\Cl}, in which there are no spin
state transitions present and in which there is only one kind of Co$^{3+}$ ion
coordination \cite{Loureiro00}. Important is that this coordination is
identical to the pyramidal CoO$_{5}$ present in the heavily debated layered
perovskites \cite{Martin97,Maignan99,Yamaura99,Loureiro00,Loureiro01,
Vogt00,Suard00,Fauth01,Burley03,Mitchell03,Moritomo00,Respaud01,Kusuya01,
Frontera02,Fauth02,Taskin03,Soda03}. Using a \textit{spectroscopic} tool, that
is soft x-ray absorption spectroscopy (XAS), we demonstrate that pyramidal
Co$^{3+}$ ions are not in the often claimed IS state but unambiguously in a HS
state. This outcome suggests that the spin states and their temperature
dependence in layered cobalt perovskites may be rather different in nature from
those proposed in the recent literature.

Bulk polycrystalline samples of {\Cl} were prepared by a solid state reaction
route \cite{Loureiro00}. The magnetic susceptibility is measured to be very
similar to the one reported by Loureiro \textit{et al.} \cite{Loureiro00}. We
find that up to 600 K the susceptibility does not follow a Curie-Weiss
behavior, making a simple determination of the spin state impossible.
Spectroscopic measurements were carried out using soft x-rays in the vicinity
of the Co-$L_{2,3}$ ($h\nu$ $\approx$ 780-800 eV) and O-$K$ ($h\nu$ $\approx$
528-535 eV) absorption edges. The experiments were performed at the Dragon
beamline at the NSRRC in Taiwan, with a photon energy resolution of about 0.30
eV and 0.15 eV, respectively. Clean sample surfaces were obtained by scraping
\textit{in-situ} with a diamond file, in an ultra-high vacuum chamber with a
pressure in the low $10^{-9}$ mbar range. The Co-$L_{2,3}$ XAS spectra were
recorded in the total electron yield mode by measuring the sample drain
current, and the O-$K$ XAS spectra by collecting the fluorescence yield to
minimize background signal and maximize bulk sensitivity. A single crystal of
EuCoO$_{3}$ is included as an unambiguous reference for a LS Co$^{III}$ system
\cite{Lorenz}.

\begin{figure}[h]
\includegraphics[width=0.5\textwidth]{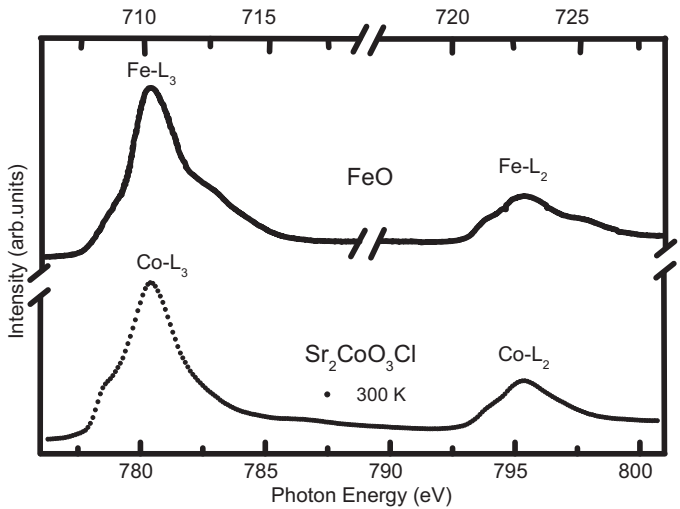}
\caption{Co-$L_{2,3}$ XAS spectrum of {\Cl} measured at 300 K ($\bullet$) and
Fe-$L_{2,3}$ XAS spectrum of FeO reproduced from Ref. 40 (solid line).}
\end{figure}\

Fig. 1 shows the Co-$L_{2,3}$ XAS spectrum of {\Cl} taken at room temperature.
It is dominated by the Co $2p$ core-hole spin-orbit coupling which splits the
spectrum roughly in two parts, namely the $L_{3}$ ($h\nu \approx 780$ eV) and
$L_{2}$ ($h\nu \approx 796$ eV) white lines regions. The line shape of the
spectrum depends strongly on the multiplet structure given by the Co $3d$-$3d$
and $2p$-$3d$ Coulomb and exchange interactions, as well as by the local
crystal fields and the hybridization with the O $2p$ ligands. Unique to soft
x-ray absorption is that the dipole selection rules are very effective in
determining which of the $2p^{5}3d^{n+1}$ final states can be reached and with
what intensity, starting from a particular $2p^{6}3d^{n}$ initial state ($n$=6
for Co$^{3+}$) \cite{deGroot94,Thole97}. This makes the technique extremely
sensitive to the symmetry of the initial state, i.e. the valence \cite{Chen90},
orbital \cite{Chen92,Park00} and spin \cite{Laan88,Thole88,Cartier92,Pen97}
state of the ion.

Utilizing this sensitivity, we compare the Co-$L_{2,3}$ XAS spectrum of {\Cl}
to that of another $3d^{6}$ compound, namely FeO, reproduced from the thesis of
J.-H. Park \cite{Parkthesis}. This spectrum was taken at room temperature.
Except for the different photon energy scale and the smaller $2p$ core-hole
spin-orbit splitting, the FeO spectrum as shown in Fig. 1 is essentially
identical with that of {\Cl}. From this we can immediately conclude that the
Co$^{3+}$ ions in {\Cl} are in the HS state, since the Fe$^{2+}$ ions are also
unambiguously HS.

To find further support for our conclusion, we also compare the Co-$L_{2,3}$
XAS spectrum of {\Cl} with that of EuCoO$_{3}$, which is known to be a LS
system \cite{Lorenz}. From Fig. 2 one now can clearly see large discrepancies
between the spectra of the two compounds. Not only are the line shapes
different, but also the ratios of the integrated intensities of the $L_{3}$ and
$L_{2}$ regions: in comparison with {\Cl}, the LS EuCoO$_{3}$ has relatively
less intensity at the $L_{3}$ and more at the $L_{2}$, characteristic for a
spin state difference \cite{Laan88,Thole88,Cartier92,Pen97}. Fig. 2 thus
demonstrates that {\Cl} is definitely not a LS system.

\begin{figure}[h]
\includegraphics[angle=0,width=0.44\textwidth]{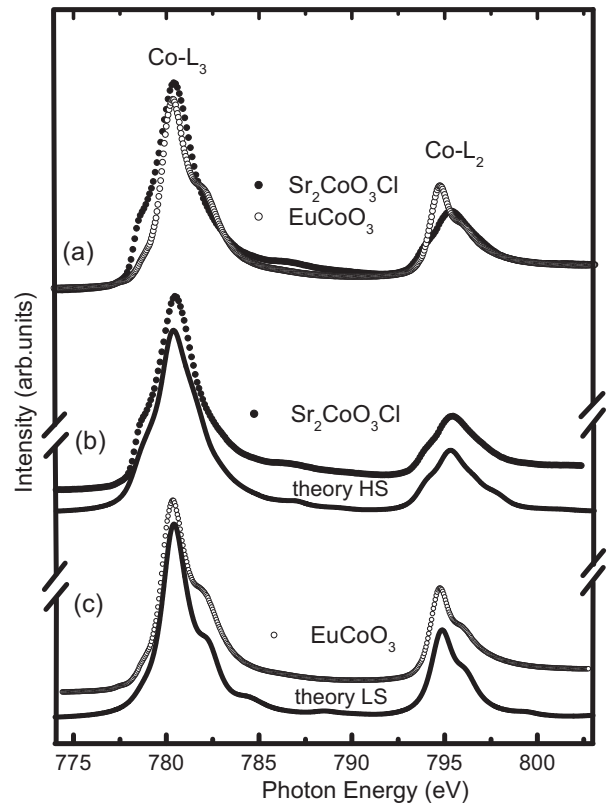}
\caption{(a) Co-$L_{2,3}$ XAS spectra of {\Cl} ($\bullet$) and EuCoO$_{3}$
($\circ$); (b) Comparison between the {\Cl } spectrum ($\bullet$) with a
theoretical simulation for a high-spin (HS) CoO$_{5}$ pyramidal cluster (solid
line); (c) Comparison between the EuCoO$_{3}$ spectrum ($\circ$) with a
theoretical simulation for a low-spin (LS) CoO$_{6}$ octahedral cluster (solid
line).}
\end{figure}\

It would have made our case even easier to prove, if we could have excluded
experimentally the IS scenario for {\Cl} by comparing the spectrum to that of a
known Co$^{3+}$ IS reference system. However, there is to date no consensus for
such an oxide reference system. Nevertheless, the spin state can also be
deduced from theoretical simulations of the experimental spectra. To this end,
we use the successful configuration interaction cluster model that includes the
full atomic multiplet theory and the hybridization with the O $2p$ ligands
\cite{deGroot94,Thole97,Tanaka94}. We have carried out the calculations for a
Co$^{3+}$ ion in the CoO$_{5}$ pyramidal cluster as present in {\Cl} and for
the ion in the CoO$_{6}$ octahedral cluster found in EuCoO$_{3}$. We use
parameter values typical for a Co$^{3+}$ system \cite{Saitoh97}. The Co $3d$ to
O $2p$ transfer integrals are adapted for the various Co-O bond lengths
according to Harrison's prescription \cite{Harrison,param}. This together with
the crystal field parameters determines whether the Co$^{3+}$ ion is in the HS
or LS state \cite{Sugano}. The results are shown in Fig. 2 and one can clearly
see that the calculated spectrum of the HS pyramidal CoO$_{5}$ cluster
reproduces very well the experimental {\Cl} spectrum, and that the calculated
LS octahedral CoO$_{6}$ spectrum matches nicely the experimental EuCoO$_{3}$
spectrum. This demonstrates that our spectroscopic assignments are firmly
founded.

\begin{figure}[h]
\includegraphics[width=0.4\textwidth]{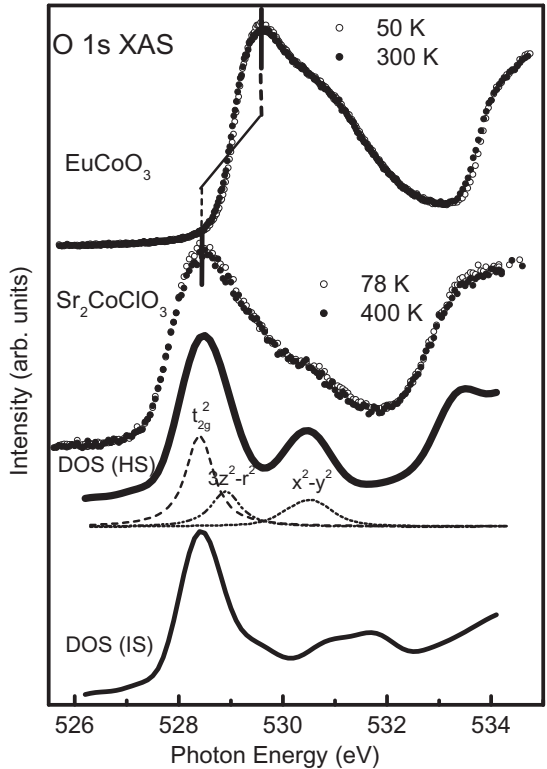}
\caption{O-$K$ XAS spectra of {\Cl} taken at 78 K ($\circ$) and 400 K
($\bullet$), and of EuCoO$_{3}$ at 50 K ($\circ$) and 300 K ($\bullet$). The
solid lines below the experimental curves depict the LDA+U calculated
unoccupied O $2p$ partial DOS for {\Cl} in the real crystal structure with the
HS state (upper) and in the artificial structure with the IS state (lower). The
dashed, dashed-dotted, and dotted lines are the $t_{2g}^{2}$, $3z^{2}$-$r^{2}$,
and $x^{2}$-$y^{2}$ projections, respectively.}
\end{figure}\

More spectroscopic evidence for the HS nature of the Co$^{3+}$ in the pyramidal
CoO$_{5}$ coordination can be found from the O $K$ XAS spectrum as shown in
Fig. 3. The structures from 528 to 533 eV are due to transitions from the O
$1s$ core level to the O $2p$ orbitals that are mixed into the unoccupied Co
$3d$ $t_{2g}$ and $e_{g}$ states. The broad structures above 533 eV are due to
Sr $4d$, Co $4s$ and Cl $3p$ related bands. For comparison, Fig. 3 also
includes the spectrum of the LS EuCoO$_{3}$, and clear differences can be seen
in the line shapes and energy positions of the Co $3d$ - O $2p$ derived states.
This again is indicative that {\Cl} is not a LS system. To interpret the
spectra, we also have carried out full-potential band structure calculations
\cite{Wu00} for {\Cl} in the local density approximation with correction for
electron correlation effects (LDA+U) \cite{Anisimov91}. We find the ground
state of the system to be an antiferromagnetic insulator with a band gap of 1.3
eV and a magnetic moment of 3.2 $\mu_B$. Although less than $4 \mu_B$, this
indicates that the Co is in the HS state since in an antiferromagnet the moment
is reduced due to covalency. The calculated unoccupied O $2p$ partial density
of states (DOS) are depicted in Fig. 3, and good agreement with the
experimental spectrum can be observed.

It is now interesting to look with more detail into the character of the states
relevant for the O $K$ XAS spectra. For the LS EuCoO$_{3}$ with the $3d$
$t_{2g}^{6}$ configuration, the lowest energy structure in the spectrum at
about 529.5 eV is due to transitions into the unoccupied Co $3d$ $e_{g}$
states. The fact that {\Cl} has a lower energy structure thus indicates that
transitions to the lower lying $t_{2g}$ are allowed, i.e. that the $t_{2g}$
states are not fully occupied. In other words, {\Cl} is in the HS
$t_{2g}^{4}e_{g}^{2}$ or IS $t_{2g}^{5}e_{g}^{1}$ state. At first sight, one
might then expect a much larger spectral weight for the higher lying $e_{g}$
level, since the hybridization with the O $2p$ is larger for the $e_{g}$ than
for the $t_{2g}$. However, our LDA+U calculations in which we find the HS
ground state, indicate that, because of the missing apical oxygen in the
CoO$_{5}$ coordination, the unoccupied $3z^{2}$-$r^{2}$ level is pulled down by
1.6 eV from the $x^{2}$-$y^{2}$, and comes close to the unoccupied
$t_{2g}^{2}$. Moreover, because of the large displacement (0.33 \AA) of the Co
ion out of the O$_{4}$ basal plane of the pyramid \cite{Loureiro00}, the
hybridization of the $x^{2}$-$y^{2}$ with the O $2p$ ligands is strongly
reduced. Therefore, the dominant lower energy structure at 528.3 eV consists of
the unoccupied minority $t_{2g}^{2}$ (dashed line in Fig. 3) and minority
$3z^{2}$-$r^{2}$ (dashed dotted line) levels, and the shoulder at 530.4 eV of
the minority $x^{2}$-$y^{2}$ (dotted line).

From the LDA+U calculations, we have found that the IS state \cite{Kwon00} is
unstable with respect to HS ground state for the real crystal structure of
{\Cl}. We have also found nevertheless, that the IS state \textit{can} be
stabilized by \textit{artificially} moving the Co ion into the O$_{4}$ basal
plane of the CoO$_{5}$ pyramid. For the latter, however, the calculated
unoccupied O $2p$ partial DOS does not reproduce the experimental O-$K$ XAS
spectrum that well, as one can see from the discrepancies in the 531-532 eV
range in Fig. 3. What happens is that the $x^{2}$-$y^{2}$ level is pushed up by
the increased hybridization with the O $2p$ ligands, since the Co ion is within
the O$_{4}$ basal plane in this artificial crystal structure. Moreover, the
up-rising majority $x^{2}$-$y^{2}$ becomes unoccupied, resulting in the IS
state. Apparently, the actual large base corrugation of the CoO$_{5}$ pyramid
helps to stabilize the HS state \cite{Wu00}, a trend that should not be
overlooked if one is to understand the real spin state of CoO$_{5}$ pyramids.
We find from our LDA+U calculations that the HS is more stable than the IS for
out-of-basal-plane Co displacements larger than a critical value of about 0.15
\AA.

Having established that the pyramidal coordinated Co$^{3+}$ ions in {\Cl} are
in the HS state, we now turn our attention to other layered cobalt materials
that have the same structural units. Neutron diffraction experiments on
$R$BaCo$_{2}$O$_{5.0}$ ($R$ = rare earth) have revealed the existence of
alternating Co$^{2+}$ and Co$^{3+}$ ions, both in pyramidal CoO$_{5}$
coordination. The magnetic structure is $G$-type antiferro with moments of 2.7
and 4.2 $\mu_{B}$ \cite{Vogt00}, or 2.7 and 3.7 $\mu_{B}$, respectively
\cite{Suard00,Fauth01}. For the $R$=Nd compound, charge ordering was not
observed, but an average moment of 3.5 $\mu_{B}$ was measured
\cite{Burley03,Mitchell03}. These studies suggested two possible scenarios for
the Co$^{3+}$ ions, namely either HS with spin-only moments or IS with orbital
moment. Our findings based on {\Cl} on the other hand, strongly suggest the HS
state of such pyramidal Co$^{3+}$ ions. Here we keep in mind that the
out-of-plane Co displacements of the pyramids in $R$BaCo$_{2}$O$_{5.0}$ are
larger than 0.35 \AA~ \cite{Vogt00,Fauth01,Burley03}, i.e. much larger than the
above mentioned 0.15 \AA~ critical value. The first scenario is thus favored,
with the remark that neutron diffraction techniques tend to observe smaller
magnetic moments due to the Co-O covalency, which is responsible for the
antiferromagnetic superexchange interactions present in these materials.

The experimental situation for the $R$BaCo$_{2}$O$_{5.5}$ system is more
complicated. Neutron and x-ray diffraction measurements indicate the presence
of all Co$^{3+}$ ions in alternating pyramidal CoO$_{5}$ and octahedral
CoO$_{6}$ units \cite{Burley03,Mitchell03,Moritomo00,Respaud01,Kusuya01,
Frontera02,Fauth02,Soda03}. The magnetic structure is most likely not a simple
$G$-type \cite{Fauth02,Taskin03}, and depending on the model, values between
0.7 and 2.0 $\mu_{B}$ have been extracted for the pyramidal Co$^{3+}$
\cite{Fauth02,Soda03}. The IS state is thus proposed, and in fact most other
studies also assumes this starting point
\cite{Burley03,Mitchell03,Moritomo00,Respaud01,Kusuya01, Frontera02,Taskin03}.
Nevertheless, structural data indicate that the CoO$_{5}$ pyramids in these
compounds have very similar Co-O bond lengths and angles as in {\Cl}. The
out-of-plane Co$^{3+}$ displacements in the pyramids are larger than 0.3 \AA~
\cite{Burley03,Respaud01,Kusuya01,Frontera02}, and again, much larger than the
0.15 \AA~ critical value. We therefore infer that also in these compounds the
pyramidal Co$^{3+}$ must be HS, which is supported by the observation that the
effective magnetic moment as extracted from the high temperature Curie-Weiss
behavior indicates a HS state for all Co$^{3+}$
\cite{Martin97,Maignan99,Moritomo00}. In fact, the average Co-O bond length for
the CoO$_{5}$ pyramids even increases at lower temperatures
\cite{Kusuya01,Frontera02}, thereby stabilizing the HS state even more. The
fact that neutron diffraction detects lower moments may indicate a complex
magnetic structure as a result of a delicately balanced spin state of the
octahedral Co$^{3+}$ ions affecting the various exchange interactions in the
compounds in which the pyramidal Co$^{3+}$ remains HS.

Summarizing, we have found an overwhelming amount of evidence for the HS nature
of the pyramidal coordinated Co$^{3+}$ ions in {\Cl}: (1) the Co $L_{2,3}$
spectrum has essentially an identical line shape as the Fe $L_{2,3}$ in FeO;
(2) the Co $L_{2,3}$ spectrum can be reproduced to a great detail by model
calculations with the Co ion in the HS state; (3) the O $K$ spectrum can be
well explained by LDA+U calculations with the Co in the HS state, but not with
the Co in the IS state; and (4) LDA+U calculations yield the HS ground state
and no stable IS state for the real crystal structure. With other newly
synthesized layered cobalt oxides having very similar pyramidal CoO$_{5}$
units, we infer that those Co$^{3+}$ ions must also be in the HS state,
contradicting the assignments made so far. It is highly desirable to
investigate the consequences for the modeling of the properties of these new
materials.

We would like to thank Lucie Hamdan for her skillful technical and
organizational assistance in preparing the experiment, and Daniel Khomskii for
stimulating discussions. The research in K\"oln is supported by the Deutsche
Forschungsgemeinschaft through SFB 608.


\begin{thebibliography}{30}
\bibitem{Briceno95} G. Briceno \textit{et al.},
Science {\bf 270}, 273 (1995).
\bibitem{Martin97} C. Martin \textit{et al.},
Appl. Phys. Lett. {\bf 71}, 1421 (1997).
\bibitem{Maignan99} A. Maignan \textit{et al.},
J. Solid State Chem. {\bf 142}, 247 (1999).
\bibitem{Takada03} K. Takada \textit{et al.},
Nature {\bf 422}, 53 (2003).
\bibitem{Fjellvag96} H. Fjellvag \textit{et al.},
J. Solid State Chem. {\bf 124}, 190 (1996).
\bibitem{Aasland97} S. Aasland \textit{et al.},
Solid State Commun. {\bf 101}, 187 (1997).
\bibitem{Kageyama97} H. Kageyama \textit{et al.},
J. Phys. Soc. Jpn. {\bf 66}, 1607 (1997).
\bibitem{Yamaura99} K. Yamaura \textit{et al.},
J. Solid State Chem. {\bf 146}, 277 (1999); Phys. Rev. B {\bf 60}, 9623 (1999);
\textit{ibid.} {\bf 63}, 064401 (2001).
\bibitem{Loureiro00} S. M. Loureiro \textit{et al.},
Chem. Mater. {\bf 12}, 3181 (2000).
\bibitem{Loureiro01} S. M. Loureiro \textit{et al.},
Phys. Rev. B {\bf 63}, 094109 (2001).
\bibitem{Vogt00} T. Vogt \textit{et al.},
Phys. Rev. Lett. \textbf{84}, 2969 (2000).
\bibitem{Suard00} E. Suard \textit{et al},
Phys. Rev. B {\bf 61}, 11871 (2000).
\bibitem{Fauth01} F. Fauth \textit{et al.},
Eur. Phys. J. B \textbf{21}, 163 (2001).
\bibitem{Burley03} J. C. Burley \textit{et al.},
J. Solid State Chem. {\bf 170}, 339 (2003).
\bibitem{Mitchell03} J. F. Mitchell \textit{et al.},
J. Appl. Phys. {\bf 93}, 7364 (2003).
\bibitem{Moritomo00} Y. Moritomo \textit{et al.},
Phys. Rev. B \textbf{61}, 13325 (2000).
\bibitem{Respaud01} M. Respaud \textit{et al.},
 Phys. Rev. B {\bf 64}, 214401 (2001).
\bibitem{Kusuya01} H. Kusuya \textit{et al.},
J. Phys. Soc. Jpn. \textbf{70}, 3577 (2001).
\bibitem{Frontera02} C. Frontera \textit{et al.},
Phys. Rev. B \textbf{65}, 180405 (2002).
\bibitem{Fauth02} F. Fauth \textit{et al.},
Phys. Rev. B \textbf{66}, 184421 (2002).
\bibitem{Taskin03} A. A. Taskin \textit{et al},
Phys. Rev. Lett. {\bf 90}, 227201 (2003).
\bibitem{Soda03} M. Soda \textit{et al.},
J. Phys. Soc. Jpn. {\bf 72}, 1729 (2003).
\bibitem{Sugano} S. Sugano, Y. Tanabe, and H. Kamimura,
\textit{Multiplets of Transition-Metal Ions in Crystals} (Academic, New York,
1970).
\bibitem{Goodenough65} J. B. Goodenough and P. M. Raccah,
J. Appl. Phys. Suppl. {\bf 36}, 1031 (1965); P. M. Raccah and J. B. Goodenough,
Phys. Rev. {\bf 155}, 932 (1967).
\bibitem{Korotin96} M. Korotin \textit{et al.},
Phys. Rev. B {\bf 54}, 5309 (1996).
\bibitem{Saitoh97} T. Saitoh \textit{et al.},
Phys. Rev. B {\bf 55}, 4257 (1997).
\bibitem{Wu00} H. Wu,
Phys. Rev. B {\bf 62}, R11953 (2000); Phys. Rev. B {\bf 64}, 092413 (2001);
Eur. Phys. J. B {\bf 30}, 501 (2002); J. Phys.: Condens. Matter {\bf 15}, 503
(2003).
\bibitem{Kwon00} S. K. Kwon \textit{et al.},
Phys. Rev. B {\bf 62}, 14637 (2000).
\bibitem{Wang01} J. Wang \textit{et al.},
Phys. Rev. B {\bf 64}, 064418 (2001).
\bibitem{Lorenz} J. Baier, diplom thesis, Univ. of Cologne (2003).
\bibitem{deGroot94} See review by F. M. F. de Groot,
J. Electron Spectrosc. Relat. Phenom. {\bf 67}, 529 (1994).
\bibitem{Thole97} See review in the Theo Thole Memorial Issue,
J. Electron Spectrosc. Relat. Phenom. {\bf 86}, 1 (1997).
\bibitem{Chen90} C. T. Chen and F. Sette,
Phys. Scr. {\bf T31}, 119 (1990).
\bibitem{Chen92} C. T. Chen \textit{et al.},
Phys. Rev. Lett. {\bf 68}, 2543 (1992).
\bibitem{Park00} J.-H. Park \textit{et al.},
Phys. Rev. B {\bf 61}, 11506 (2000).
\bibitem{Laan88} G. van der Laan \textit{et al.},
Phys. Rev. B {\bf 37}, 6587 (1988).
\bibitem{Thole88} B. T. Thole and G. van der Laan,
Phys. Rev. B \textbf{38}, 3158 (1988).
\bibitem{Cartier92} C. Cartier dit Moulin \textit{et al.},
J. Phys. Chem. {\bf 96}, 6196 (1992).
\bibitem{Pen97} H. F. Pen \textit{et al.},
Phys. Rev. B {\bf 55}, 15500 (1997).
\bibitem{Parkthesis} J.-H. Park, PhD Thesis,
University of Michgan (1994).
\bibitem{Tanaka94} A. Tanaka and T. Jo,
J. Phys. Soc. Jpn. \textbf{63}, 2788 (1994).
\bibitem{Harrison} W. A. Harrison,
\textit{Electronic Structure and the Properties of Solids} (Dover, New York,
1989).
\bibitem{param} Parameters for HS CoO$_{5}$: $\Delta$=2.5, $U_{dd}$=5.5, $U_{\underline{c}d}$=7.0,
$V_{b1}$=2.4, $V_{a1}$=2.0, $V_{b2}$=1.21, $V_{e}$=1.23; $10Dq$=0.9, $Ds$=0.06,
$Dt$=0.05, $T_{pp}$=0.3 (eV); Parameters for LS CoO$_{6}$: $\Delta$=2.5,
$U_{dd}$=5.5, $U_{\underline{c}d}$=7.0, $V_{e_{g}}$=2.6, $V_{t_{2g}}$=1.38,
$10Dq$=1.0, $T_{pp}$=0.5 (eV).
\bibitem{Anisimov91} V.I. Anisimov \textit{et al.},
Phys. Rev. B \textbf{44}, 943 (1991).
\end{thebibliography}
\end{document}